\documentclass[%
 reprint,
 amsmath,amssymb,
 aps,
 pre,
 superscriptaddress
]{revtex4-1}

\usepackage{CJK}
\usepackage{graphicx}
\usepackage{dcolumn}
\usepackage{bm}
\usepackage{xcolor}
\usepackage{amsmath}
\usepackage{amsfonts}
\usepackage{mathtools}
\usepackage{textcomp}
\usepackage{pdfpages}
\usepackage{pgffor}

\makeatletter
\AtBeginDocument{\let\LS@rot\@undefined}
\makeatother

\begin{document}

\begin{CJK*}{UTF8}{} 

\title{Determination of Glass Transition Temperature of Polyimides from Atomistic Molecular Dynamics Simulations and Machine-Learning Algorithms}

\author{Chengyuan Wen ({\CJKfamily{gbsn}温程远})}
\affiliation{Department of Physics and Center for Soft Matter and Biological Physics, Virginia Polytechnic Institute and State University, Blacksburg, Virginia 24061, USA}
\affiliation{Macromolecules Innovation Institute, Virginia Polytechnic Institute and State University, Blacksburg, Virginia 24061, USA}
\author{Binghan Liu ({\CJKfamily{gbsn}刘秉汉})}
\affiliation{Department of Physics and Center for Soft Matter and Biological Physics, Virginia Polytechnic Institute and State University, Blacksburg, Virginia 24061, USA}
\affiliation{Macromolecules Innovation Institute, Virginia Polytechnic Institute and State University, Blacksburg, Virginia 24061, USA}
\author{Josh Wolfgang}
\affiliation{Macromolecules Innovation Institute, Virginia Polytechnic Institute and State University, Blacksburg, Virginia 24061, USA}
\affiliation{Department of Chemistry, Virginia Polytechnic Institute and State University, Blacksburg, Virginia 24061, USA}
\author{Timothy E. Long}
\affiliation{Macromolecules Innovation Institute, Virginia Polytechnic Institute and State University, Blacksburg, Virginia 24061, USA}
\affiliation{Department of Chemistry, Virginia Polytechnic Institute and State University, Blacksburg, Virginia 24061, USA}
\author{Roy Odle}
\affiliation{SABIC Innovative Plastics, 1 Lexan Lane, Mt. Vernon, Indiana 47620, USA}
\author{Shengfeng Cheng ({\CJKfamily{gbsn}程胜峰})}
\email{chengsf@vt.edu}
\affiliation{Department of Physics and Center for Soft Matter and Biological Physics, Virginia Polytechnic Institute and State University, Blacksburg, Virginia 24061, USA}
\affiliation{Macromolecules Innovation Institute, Virginia Polytechnic Institute and State University, Blacksburg, Virginia 24061, USA}
\affiliation{Department of Mechanical Engineering, Virginia Polytechnic Institute and State University, Blacksburg, Virginia 24061, USA}

\date{\today}

\begin{abstract}
Glass transition temperature ($T_{\text{g}}$) plays an important role in controlling the mechanical and thermal properties of a polymer. Polyimides are an important category of polymers with wide applications because of their superior heat resistance and mechanical strength. The capability of predicting $T_{\text{g}}$ for a polyimide \textit{a priori} is therefore highly desirable in order to expedite the design and discovery of new polyimide polymers with targeted properties and applications. Here we explore three different approaches to either compute $T_{\text{g}}$ for a polyimide via all-atom molecular dynamics (MD) simulations or predict $T_{\text{g}}$ via a mathematical model generated by using machine-learning algorithms to analyze existing data collected from literature. Our simulations reveal that $T_{\text{g}}$ can be determined from examining the diffusion coefficient of simple gas molecules in a polyimide as a function of temperature and the results are comparable to those derived from data on polymer density versus temperature and actually closer to the available experimental data. Furthermore, the predictive model of $T_{\text{g}}$ derived with machine-learning algorithms can be used to estimate $T_{\text{g}}$ successfully within an uncertainty of about 20 degrees, even for polyimides yet to be synthesized experimentally.
\end{abstract}

\maketitle

\end{CJK*}

\section{Introduction}\label{sec:Intro}

When a polymer is rapidly cooled below a certain temperature, it undergoes a transition into a glassy phase where the polymer has an amorphous structure but exhibits rigidity on experimental time scales. The temperature at which this transition occurs is termed the glass transition temperature ($T_\text{g}$) and is one of the most important properties of a polymer that determine its performance and applications. For example, if a polymer has to stay as a hard solid in a certain application, its $T_\text{g}$ should be much higher than the environmental temperature, $T_\text{e}$. On the other hand, if a rubber or a polymer melt is required, then $T_\text{g}$ needs to be lower than $T_\text{e}$. The difference between $T_\text{g}$ and $T_\text{e}$ also strongly affects other physical properties of the polymer, such as its density and the diffusion behavior of guest gas molecules in the polymer. In other words, many physical properties of a polymer exhibit changes, which can be significant, when $T_{\text{e}}$ is varied to cross $T_\text{g}$. This observation is underlying a variety of methods that are used to determine $T_\text{g}$ via measuring or computing these physical properties as a function of temperature. The glass transition temperature is therefore a critical parameter to be considered when the target is to design or identify a polymeric material that meets the requirement of a given application. In this paper, we explore three approaches to determine $T_\text{g}$ for various polyimides \textit{in silico}, including computing their density and gas diffusion coefficients in the polymers with atomistic molecular dynamics (MD) simulations and predicting $T_\text{g}$ with a model derived by applying machine-learning algorithms to analyze the existing structure-property data on $T_\text{g}$ of polyimides collected from literature.

Polyimides are a category of engineering plastics that have wide applications in the automotive and aerospace industries because of their relatively high $T_\text{g}$, high strength, and good heat resistance properties.\cite{wilson1990polyimides,mittal1984polyimides} Experimentally, $T_\text{g}$ can be measured with differential scanning calorimetry (DSC) and thermo-mechanical analysis techniques. However, these procedures usually require careful sample preparation and control of the measurement conditions. As a supplementary approach to expedite material characterization, MD simulations have been used to quantify $T_\text{g}$ since the 1980s. Rigby \textit{et al.} calculated $T_\text{g}$ for Kremer-Grest chains consisting of Lennard-Jones beads as a model of polyethylene.\cite{rigby1987molecular} The temperature dependence of the polymer density, the self-diffusion coefficient, and the characteristic ratio $\langle r^2 \rangle/(nl^2)$ was used to identify the glass transition and $T_\text{g}$. Han \textit{et al.} calculated $T_\text{g}$ for five different polymers using the curve of specific volume against temperature.\cite{han1994glass} Abu-Sharkh \textit{et al.} used a similar method to compute $T_\text{g}$ for poly(vinylchloride)s with the force field determined with an \textit{ab initio} method.\cite{abu2001glass} In these studies, usually only one polymer chain was simulated for each system because of the limitation of computational power. Morita \textit{et al.} simulated 100 coarse-grained polymer chains and introduced a method to compute $T_\text{g}$ by examining the mean-square displacement of a polymer segment at different temperatures.\cite{Morita2006Study} Buchholz \textit{et al.} studied the cooling-rate-dependence of $T_\text{g}$ with the Kremer-Grest model, where $T_\text{g}$ was found from the curves of nonbonded energy or system volume versus temperature.\cite{buchholz2002cooling} Following these initial efforts, other researchers started to compute $T_\text{g}$ with atomistic MD simulations for various polymers,\cite{fan1997local, deazle1996molecular, Hamerton2006Developing, pozuelo2002glass, hu2006molecular, wang2016study} typically by investigating the density change of a polymer when the temperature is varied. Lyulin \textit{et al.} calculated the $T_\text{g}$ of several polyimide polymers and pointed out that the results from MD simulations depend on cooling rate and vary if the atomistic model of a polymer considers partial charges or not.\cite{lyulin2014thermal} Mohammadi \textit{et al.} computed $T_\text{g}$ of poly(methyl methacrylate) using the first peak of the pair correlation function, the mean square displacement of polymer segments, the self-diffusion coefficient, and the internal energy of the system.\cite{mohammadi2017glass} They found that though all these different physical quantities exhibit an obvious transition at $T_\text{g}$, the values from MD simulations are usually lower than the experimental value. All the reported work thus shows that MD simulations can be a useful tool to obtain $T_\text{g}$ but the results can suffer from small system sizes, short chain lengths, and high cooling rates, all reflecting the limitations of MD methods. Furthermore, the results may depend on the particular force field being used in a study.\cite{sun2018molecular,Alzate_Vargas_2018,luchinsky2018molecular}

Although MD methods can be used to calculate $T_\text{g}$ for a polymer, the calculations can still take a long time and may be limited by available computational resources. Therefore, a predictive model of $T_\text{g}$ is highly desirable, which uses certain features of a polymer, such as the chemical identity of the monomer and the sequence structure of the chain, as inputs. Such a model can be applied to quickly yield $T_\text{g}$ that can be tested later with MD calculations or experiments. This capability will allow quick screening of a series of polymers when a particular application is in consideration. Efforts of generating so-called quantitative structure property relationships (QSPRs) for the glass transition temperature of a polymer have been ongoing since the 1990s.\cite{sumpter1994neural,joyce1995neural} Joyce \textit{et al.} used neural network algorithms to train a model for $T_\text{g}$ prediction with data on 360 monomers and the model can predict $T_\text{g}$ for other 89 monomers with a root mean square error (RMSE) of about 35 K.\cite{joyce1995neural} The large error may be caused by the fact that the 360 monomers picked by Joyce \textit{et al.} were for a broad range of polymers and with a small dataset, neural network algorithms could easily lead to overfitting. Yu \textit{et al.} applied the multiple linear stepwise regression method to establish a predictive model of $T_\text{g}$ with a RMSE around 15.2 K.\cite{yu2006prediction}

Chen \textit{et al.}\cite{chen2008neural}, Ning \textit{et al.}\cite{ning2009artificial}, and Xu \textit{et al.}\cite{xu2012prediction} also developed predictive models of $T_\text{g}$ with different accuracy for a variety of polymers. The number of data points in their training set ranges from 52 to 107. Pei \textit{et al.} applied support vector regression (SVR) optimized by an integrated particle swarm optimization to predict $T_\text{g}$.\cite{pei2013modeling} They used 25 sample points to train the model and 7 other data points to test it. The RMSE of their model is around 4 K. However, the penalty factor of 56700186162.908470 used in their model training procedure is not replicable. Chen \textit{et al.} applied multiple linear regression analysis to establish a predictive model of $T_\text{g}$ with 60 training data points.\cite{chen2018computational} The test set contained 20 data points and the prediction error of $T_\text{g}$ was around 58 K.

Despite the existing efforts of constructing predictive models of $T_\text{g}$ for a range of polymers, the stability of such models has not been proved or discussed. It is unclear if the models reported in literature are robust and possess the same predictive power and accuracy if the training and test datasets are split in different ways. In this paper, we discuss the instability issue of the commonly used regularization method termed ``least absolute shrinkage and selection operator'' (LASSO) and find that a bragging approach can be used to enhance the stability of the predictive model of $T_\text{g}$ derived with LASSO. Furthermore, we compare $T_\text{g}$ predicted by the model trained with machine-learning algorithms with those computed with atomistic MD simulations for several polyimides that were yet synthesized at the time of prediction and computation. Later on, these polyimides were synthesized and their glass transition temperatures were measured using DSC. The predicted and computed values agree with the experimental results within 10 to 20 K. This comparison not only serves as a test of the predictive model but also validates the power of MD methods of computing $T_\text{g}$ for polyimides with new formulae.

This paper is organized as follows. In Sec.~\ref{sec:ch3_md}, the methods of determining $T_\text{g}$ with atomistic MD simulations are introduced and the results are analyzed and discussed. Then in Sec.~\ref{sec:ch3_ml}, the procedure of building a predictive model of $T_\text{g}$ for polyimides using machine-learning algorithms is discussed in detail, including dataset preparation and separation, the digitization of polymer structures, the conversion of polymer structures to proper SMILES notations, the generation of polymer features from their SMILES notations, and the construction of the predictive model (i.e., the mapping from polymer features to $T_\text{g}$) via machine learning. Finally, conclusions are presented in Sec.~\ref{sec:ch3_conclusion}.

\section{Determination of Glass Transition Temperature with All-Atom Molecular Dynamics Simulations}\label{sec:ch3_md}

\subsection{All-Atom Molecular Dynamics Simulation Methods}

\begin{figure*}[htb]
  \centering
  \includegraphics[width=\textwidth]{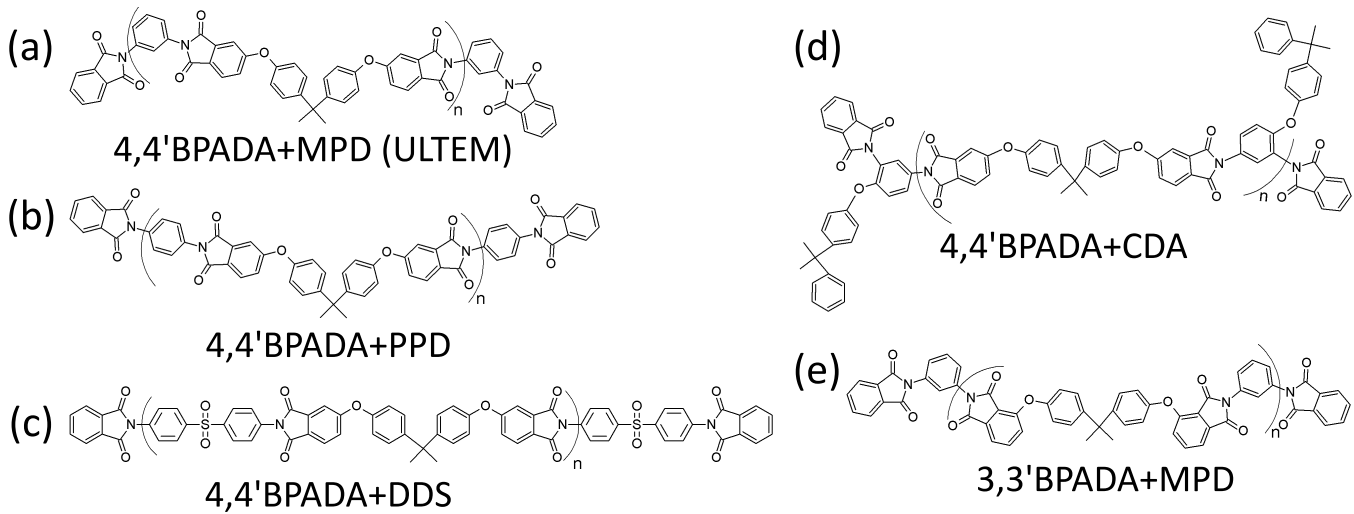}
  \caption{Structures of polyetherimides studied with all-atom MD simulations.}
  \label{fig:chem_form}
\end{figure*}

Atomistic MD simulations were employed previously to model the mechanical, thermal, and dielectric properties of polyetherimides.\cite{falkovich2014influence, xia2010structural, luchinsky2018molecular} The polyimide chains in our study were built with MAPS builder.\cite{maps} In this section, 5 polyetherimides were studied, including 4,4'BPADA+MPD (ULTEM), 4,4'BPADA+PPD, 3,3'BPADA+MPD, 4,4'BPADA+CDA, and 4,4'BPADA+DDS.\footnote{BPADA: 4,4'-bisphenol A dianhydride; MPD: m-phenylenediamine; PPD: para-phenylenediamine; CDA 1,2-dihydroxybenzene dianhydride: DDS: diphenyl sulfone.} The chemical formulae of these polymers are shown in Fig.~\ref{fig:chem_form}. Each chain consists of 4 repeating units (i.e., $n=4$ in Fig.~\ref{fig:chem_form}) and the molecular weights range from 2.7 to 3.8 kDa. Phthalic anhydride (PA) groups are added to cap the chains. All MD simulations were performed using LAMMPS\cite{Plimpton1995} with the PCFF force field.\cite{sun1994} The equations of motion were integrated with a velocity-Verlet algorithm with time step $\Delta t = 1$ fs. The Mulliken charge was included in the model and calculated using Gaussian09 software with the semi-empirical PM6 method as the basis set.\cite{g16} The cutoffs of nonbonded and Coulomb interactions were both set as 12 \AA ~and the long-range part of Coulomb interactions was calculated using the particle-particle particle-mesh method. Each system contained 512 chains. A hydrostatic pressure of 1000 atm was used to compress the system at 300 K until it reached a density around 1.2 g/$\text{cm}^{3}$, close to the experimental value of ULTEM.\cite{bashford1997polyetherimides} Then the system was heated up to 800 K under 1 atm and equilibrated at 800 K for 5 ns. After this step, the system was gradually cooled down to 300 K under 1 atm. In this process, many configurations were created for a series of temperatures between 300 K and 800 K. At a given temperature, the corresponding configuration was taken as a starting state and the system was equilibrated further for 2 ns. The density of the polymer was then computed in a NPT ensemble with a target pressure at 1 atm controlled by a Nose-Hoover barostat. The temperature was controlled with a Nose-Hoover thermostat. The equilibrated system was also used for simulating the diffusion of gas molecules in the polymer. In these simulations, a NVT ensemble was adopted.

\begin{figure}
  \centering
  \includegraphics[width=0.45\textwidth]{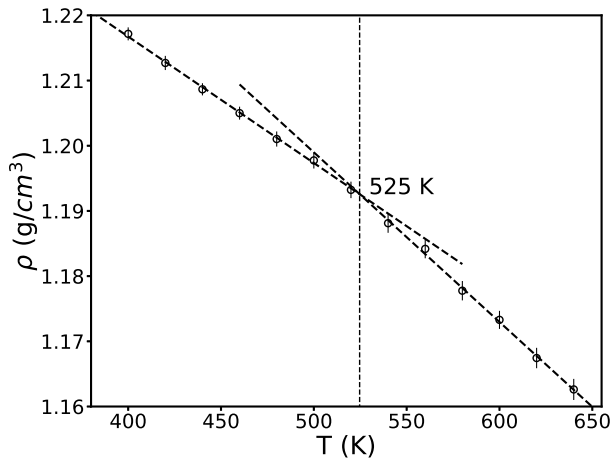}
  \caption{$\rho(T)$ vs. $T$ for 4,4'BPADA+MPD (ULTEM), for which $T_\text{g}=525$ K.}
  \label{fig:Tg_den_ULTEM}
\end{figure}

A commonly used protocol to determine $T_\text{g}$ for a polymer is to calculate its density as a function of temperature, i.e., to obtain the $\rho(T)$ curve.\cite{Yang2016, Li2009, Lyulin2014, Minelli2012, Yoshimizu2012, Yu2001} One example is shown in Fig.~\ref{fig:Tg_den_ULTEM} for ULTEM (i.e., 4,4'BPADA+MPD). The value of $T_\text{g}$ can be determined from the intersection of the two linear fits to $\rho(T)$, one for the lower and the other for the higher temperature region. At room temperature, the density of ULTEM from MD simulations is slightly lower than the experimental value, $1.27~\text{g/}\text{cm}^3$. The data indicate that the variation of density with temperature is captured by the PCFF force field. The value of $T_\text{g}$ computed from MD data on $\rho(T)$ is 525 K for ULTEM, which is 35 K higher than the experimental value, 490 K. The $\rho(T)$ curves for other polyetherimides in Fig.~\ref{fig:chem_form} are included in the Supporting Information. The results on their $T_\text{g}$ are listed in Table~\ref{tb:TgData}. The values determined using the $\rho(T)$ data are generally 20 to 30 K higher than the corresponding experimental values. 

In addition to density, there are other properties of a polymer that can be used to determine $T_\text{g}$, including volume, free volume, specific volume, radial distribution functions, mean-square displacements, nonbonded energy, dihedral torsion energy, etc.\cite{Yang2016,Li2009} Many studies also showed that the diffusion behavior of gas molecules in a polymer matrix changes when the polymer undergoes a glass transition.\cite{meares1957diffusion, kumins1961diffusion} This can be understood by examining the temperature dependence of the diffusion coefficient of a gas molecule in a polymer matrix, which has an Arrhenius form,
\begin{equation}
D=D_0~\text{exp}\left(-\frac{E_A}{RT}\right)~,
\end{equation}
where $D$ is the diffusion coefficient of the gas molecules, $D_0$ is a prefactor with the same unit as $D$, $E_A$ is the activation energy for diffusion, $T$ is the absolute temperature, and $R=8.314~\text{J~mol}^{-1}$ is the gas constant. Note that $E_A$ may be temperature dependent but in many cases the dependence is weak and negligible.\cite{menzinger1969meaning} For a glassy polymer, the value of $E_A$ changes when the polymer enters a glassy state from a melt state. Therefore, a plot of $D(T)$ vs. $1/T$ on a log-linear scale will show a straight line for $T>T_\text{g}$ and another straight line with a different slope for $T<T_\text{g}$. The intersections between these two lines can be used to determine $T_\text{g}$.

\begin{figure}[htb]
  \centering
  \includegraphics[width=0.45\textwidth]{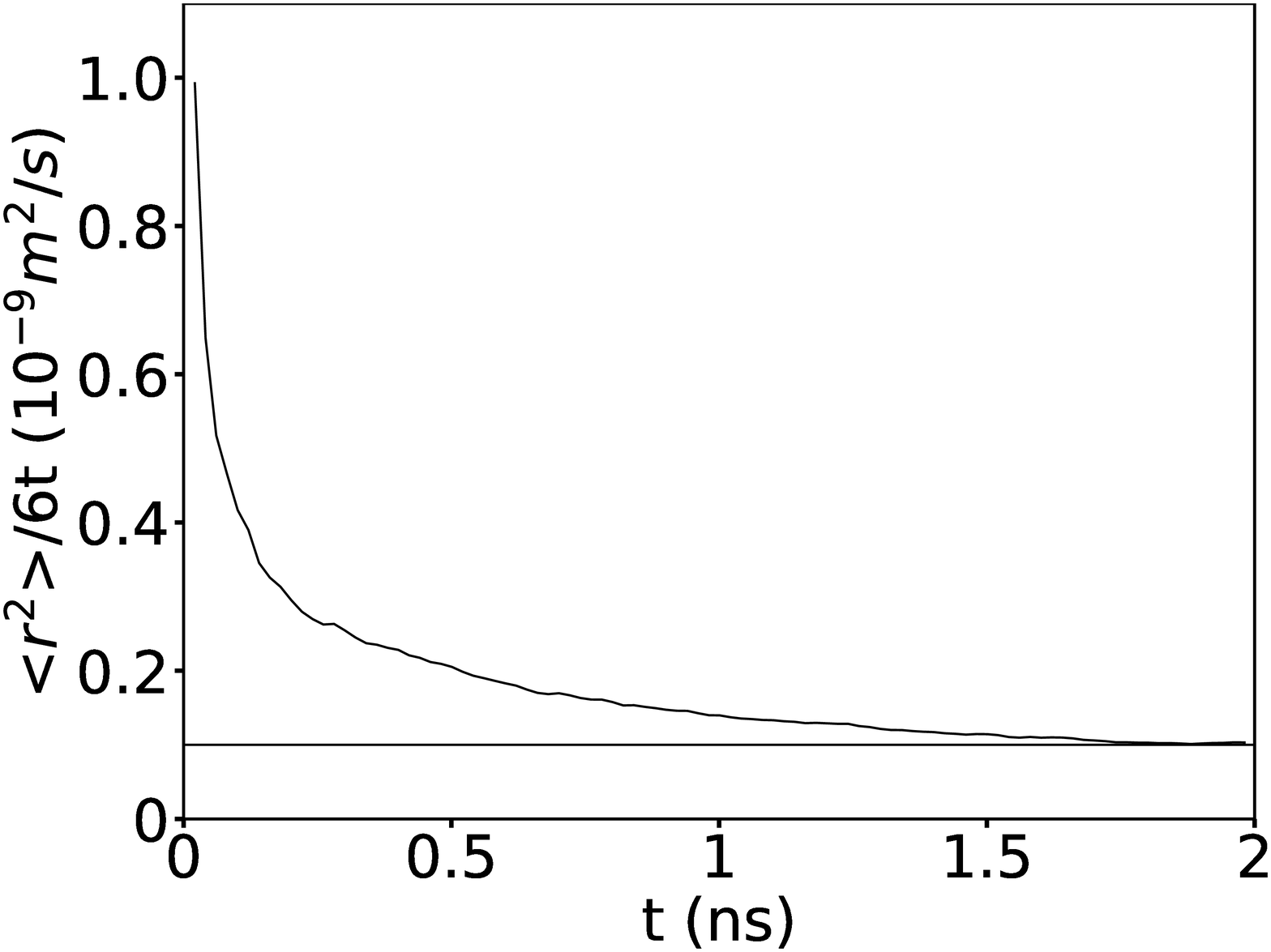}
  \caption{$\langle r^2 \rangle /(6t)$ vs. time $t$ for argon diffusing in 4,4'BPADA+MPD (ULTEM) at 400 K, which yields $D\simeq 1.0\times {\rm 10^{-10} m^2/s}$. }
  \label{fig:diff_coeff_ULTEM}
\end{figure}

\begin{figure}
  \centering
  \includegraphics[width=0.45\textwidth]{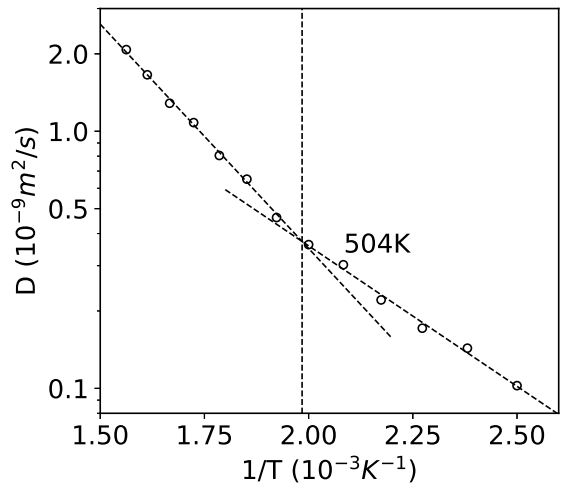}
  \caption{$D(T)$ vs. $1/T$ on a log-linear scale for argon diffusing in 4,4'BPADA+MPD (ULTEM), which yields $T_\text{g}=504$ K.}
  \label{fig:Tg_diff_ULTEM}
\end{figure}

In MD simulations, the diffusion coefficient $D$ of a gas molecule in a polymer can be computed from its mean-square displacement (MSD), $\langle r^2 \rangle$. In the diffusive regime, its time dependence can be expressed as
\begin{equation}
\langle r^2 \rangle = 6Dt + C,
\end{equation}
where $t$ is time and $C$ is a constant. In our simulations, 1000 inert gas atoms, either argon or neon, were added to the polymer system and their diffusion was tracked. The average MSD was then computed. One example is shown in Fig.~\ref{fig:diff_coeff_ULTEM}, where we plot $\langle r^2 \rangle /(6t)$ vs. $t$ for argon diffusing in 4,4'BPADA+MPD (i.e., ULTEM) at 400 K. Clearly, $\lim_{t\rightarrow \infty} \langle r^2 \rangle /(6t) = D$. This calculation can be performed at various temperatures to generate the $D(T)$ curve. Figure~\ref{fig:Tg_diff_ULTEM} shows the results for argon diffusing in ULTEM, where $D$ is plotted against $1/T$ on a log-linear scale. The two regions in which $\log D$ depends $1/T$ linearly are visible. The corresponding linear fits and their intersection are used to determine $T_\text{g}$. For ULTEM, we find that $T_\text{g} \simeq 504$ K, which compared with the result determined using the $\rho(T)$ data is much closer to the experimental value (490 K). The $D(T)$ results for other polyetherimides in Fig.~\ref{fig:chem_form} are included in the Supporting Information.

\subsection{Molecular Dynamics Simulation Results and Discussion}

\begin{table*}[htb]
\centering
\caption{Summary of $T_\text{g}$ (K) for various polyetherimides from MD calculations of $\rho(T)$ and $D(T)$, the predictive model constructed using machine-learning algorithms, and DSC measurements.}
\label{tb:TgData}
\begin{tabular}{|m{2.25cm}|m{1.95cm}|m{1.95cm}|m{1.95cm}|m{1.95cm}|m{1.95cm}|}
\hline
& {\footnotesize 4,4'BPADA + MPD} & {\footnotesize 4,4'BPADA + PPD} &{\footnotesize 4,4'BPADA + DDS} & {\footnotesize 4,4'BPADA + CDA} & {\footnotesize 3,3'BPADA + MPD}\\ \hline
{\small MD [$\rho(T)$]} & 525 & 539 & 542 & 516 & 524\\ \hline
{\small MD [$D(T)$] } & 504 & 488 & 532 & 495 & 500 \\ \hline
{\small Predicted} & 515 & 499 & 537 & 551 & 528 \\ \hline
{\small Experimental} & 490 & unknown & 520 & 473 & 511\\ \hline
\end{tabular}
\end{table*}

All results on $T_\text{g}$ determined using either $\rho(T)$ or $D(T)$ that were computed with all-atom MD simulations are summarized in Table~\ref{tb:TgData}. The predicted values of $T_\text{g}$ by the model derived with a machine-learning approach discussed in Sec.~\ref{sec:ch3_ml} and the experimental results for 4 polyetherimides measured with DSC are included as well. It must be emphasized that 3 of them, including 4,4'BPADA+DDS, 4,4'BPADA+CDA, and 3,3'BPADA+MPD, were synthesized and had their $T_\text{g}$ measured after the computation and prediction were performed. It is noted that the values of $T_\text{g}$ from the $\rho(T)$ curves are generally 20 to 30 K higher than the available corresponding experimental values. However, the results on $T_\text{g}$ from the $D(T)$ data are closer to and only about 10 to 20 K higher than the experimental $T_\text{g}$'s.

The method of using gas diffusion coefficients to determine $T_\text{g}$ has several advantages. First of all, the results on $T_\text{g}$ from $D(T)$ agree better with the experimental values, as evidenced by the data in Table~\ref{tb:TgData}. Secondly, the diffusion coefficient of a gas molecule can be computed quickly and accurately with MD simulations. Such calculations only require fairly short MD trajectories ($\sim 1$ to 2 ns). The self-diffusion coefficient of a polymer can also be used to pinpoint $T_\text{g}$. However, a polymer typically diffuses much more slowly than gas molecules. As a result, much longer MD simulations are needed to compute the self-diffusion coefficient of a polymer to the same precision as in the diffusion coefficient of gas molecules. Finally, to compute $D(T)$ we can use a NVT ensemble with temperature well controlled by a suitable thermostat (e.g., a Nose-Hoover thermostat). However, to compute the $\rho(T)$ curve, a NPT ensemble is required, which needs both a thermostat and a barostat. In MD simulations, it is practically very challenging to control pressure accurately, particularly if the pressure is as small as 1 atm.\cite{heyes1983molecular,feller1995constant} Our MD data show that when the target pressure is 1 atm, the actual pressure in the system can fluctuate significantly from about -190 atm to about 190 atm. As a result, the polymer density also fluctuates strongly and an average over a long period of time (i.e., a long MD trajectory) is required to generate a $\rho(T)$ curve with a reasonable accuracy. Computing $D(T)$ instead of $\rho(T)$ circumvents this issue and leads a much faster convergence of the data that can be used to determine $T_\text{g}$, which is especially the case in the high-temperature range.

\section{Predictive Model of Glass Transition Temperature Trained with Machine-Learning Algorithms} \label{sec:ch3_ml}

\subsection{Machine-Learning Methods}

Machine learning is considered a subset of artificial intelligence. A machine-learning algorithm is a mathematical model that can be trained by a set of sample data without requiring the system to be explicitly programmed to generate (pre-determined) outputs on the basis of given inputs. After the training process, such a mathematical model can be used to make a future decision or prediction given new data. There are three basic machine-learning paradigms: supervised, unsupervised, and reinforcement. The process adopted here is a supervised learning method as the sample dataset used for training includes both inputs (e.g., polymer chemical identity and sequence) and desired outputs (e.g., glass transition temperature). The outcome of the learning process is an optimized objective function that connects the chemical information of a polymer, particularly its monomer type and sequence, to its measurable physical property.

Many efforts have been devoted to synthesize various polyimides, characterize their structures, and measure their properties including $T_\text{g}$.\cite{fang2002polyimides,hsiao1998synthesis, takahashi1998preparation,li2003polyimides,zhang2006polyimides} By collecting available data published in literature and applying machine-learning approaches to analyze the data, we can develop a predictive model for the glass transition temperature of polyimides. This model can be used to probe polyimides that are yet to be synthesized. In particular, in this paper we will compare the predictions of $T_\text{g}$ from the mapping function derived via machine learning with those computed with atomistic MD simulations for a few selected polyimides before they are made in a lab. This comparison serves as a test of the machine-learning-generated predictive model. In the future, various polyimides with potential values in terms of their performance and application will be screened with the predictive model and then selected formulae will be synthesized in a lab to validate and improve the model.

\subsubsection{Database and Feature Generation}

We collected 225 data points on the glass transition temperature of polyimides from literature, including 160 data points from Ref.~\cite{ding2007isomeric} and 65 data points from Ref.~\cite{liu2010prediction}. Some sample data are shown in Table~\ref{tb:sample_Tg_data}. For each polymer, the chemical identity of the monomer is taken as the input. The skeleton notation of a polyimide was drawn and converted into an expression called Simplified Molecular-Input Line-Entry System (SMILES), which is a line-notation system using an ASCII string to represent the structure of a polymer. Then a feature-generating engine called E-dragon was utilized to read in the generated SMILES notations and to extract the available features for each polyimide.\cite{alvaDesc} In polymer informatics, features are also called descriptors, consisting of individual measurable properties of a molecule or a polymer.\cite{Ramprasad2017, Audus2017, Peerless2019} The ensemble of descriptors represents the characteristics of the polymer/molecule being studied. As polyimides are made of dianhydrides and diamines, we calculated the features for a dianhydride and a diamine group separately. For each polyimide, E-dragon generates 1342 descriptors for its dianhydride group and the same set for its diamine group. Sample features include molecular weight, sum of atomic van der Waals volumes, and sum of atomic polarizabilities, etc.

\begin{table*}[htb]
\footnotesize
  \begin{center}
    \caption{Sample dataset of $T_\text{g}$ of polyimides from Refs.~\cite{ding2007isomeric} and \cite{liu2010prediction}.}
    \vspace*{3mm}
    \label{tb:sample_Tg_data}
    \begin{tabular}{|c|l|l|c|} 
     \hline
      \textbf{No.} &  \textbf{Polyimide's Name} & \textbf{SMILES Notation} & \textbf{$T_\text{g}$ (K)} \\
      \hline
      1&4,4'TDPA+1,4,4APB &Nc1ccc(cc1)$\cdots$(=O)OC(=O)c7c6

 & 234 \\ 
      2&3,3'ODPA+M,M'DABP &Nc1cccc(c1)$\cdots$(Oc4cccc3C(=O)OC(=O)c34)c5C6=O
 & 234\\
      3&4,4'ODPA+M,M'DABP &Nc1cccc(c1)$\cdots$c5ccc6C(=O)OC(=O)c6c5
 & 235\\
      4&3,3'ODPA+M,M'DDS & Nc1cccc(c1)$\cdots$(Oc4cccc3C(=O)OC(=O)c34)c5C6=O
& 241\\
      5&3,4'TDPA+1,4,4APB &Nc1ccc(cc1)$\cdots$(Sc4ccc5C(=O)OC(=O)c5c4)c6C7=O
 & 242\\
      6&3,4'ODPA+M,M'DABP & Nc1cccc(c1)$\cdots$(Oc3ccc4C(=O)OC(=O)c4c3)c5C6=O
& 243\\
      7&4,4'BPDA+M,M'ODA &Nc1cccc(c1)$\cdots$c6C(=O)OC(=O)c6c5
 & 243\\
      8&4,4'BPDA+p,p'ODA
 &Nc1ccc(cc1)$\dots$OC(=O)c6c5
 & 262\\
      9&4,4'BTDA+m,m'MDA
&Nc1cccc(c1)$\dots$c5ccc6C(=O)OC(=O)c6c5
 & 272\\
      10&4,4'BTDA+o,o'MDA
 &Nc1ccc(cc1)$\dots$c7cccc6C(=O)OC(=O)c67
 & 283\\
      ...& ... & ... & ... \\
 \hline
    \end{tabular}
  \end{center}
\end{table*}

\subsubsection{Data Splitting into Training Set and Test Set}\label{subsubsec:ch3_splitting}

\begin{figure}[htb]
  \centering
  \includegraphics[width=0.45\textwidth]{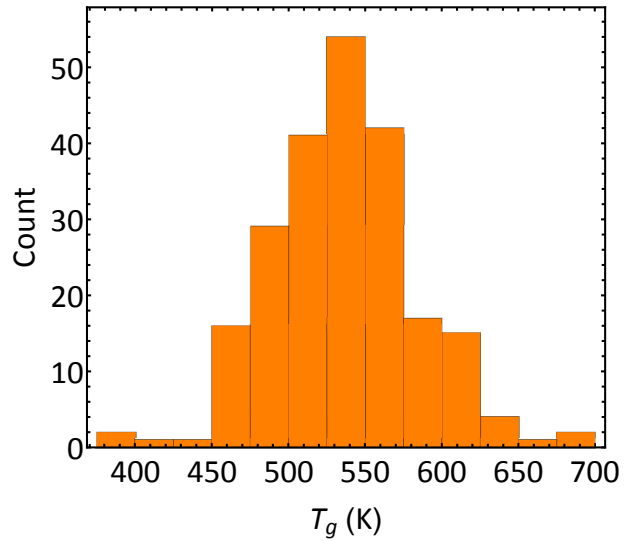}
  \caption[caption for Tg distribution figure]{The distribution of $T_\text{g}$ of polyimides collected from Refs.~\cite{ding2007isomeric} and \cite{liu2010prediction}.}
  \label{fig:Tg_dist}
\end{figure}

For polyimides, the values of $T_\text{g}$ collected from Refs.~\cite{ding2007isomeric} and \cite{liu2010prediction} range from 273 K to 697 K. However, the distribution is not uniform in this range. The majority of the data is between 466 K and 583 K. The distribution of the 225 data points on $T_\text{g}$ is shown in Fig.~\ref{fig:Tg_dist}, with a peak around 530 K. The nonuniform nature of the distribution must be considered when the dataset is split into a training set and a test set as it is important for the training set to be representative of the entire dataset. This is particularly a concern if the number of available data is limited, as in the case here. To examine the influence of how the dataset is split on the performance of the resulting predictive model of $T_\text{g}$, we test two different ways of dividing the dataset into a training and a test set. To this end, we only use the 160 data points from Ref.~\cite{ding2007isomeric} to train the model and reserve the 65 data points from Ref.~\cite{liu2010prediction} for a completely independent test of the predictive capability of the machine-learning-trained model. To ensure that the relatively small dataset can be split consistently, we first remove the data points of $T_\text{g}$ at the tail of the probability distribution, i.e., those below 423 K or above 623 K. The total number of the remove data points is 9, leaving 151 points in the dataset. In the first way, this dataset is randomly split into a training set containing 85\% of the data and a test set consisting of the remaining 15\%. In the second way, the dataset is first divided into 8 adjoining sections, each of width of 25 K. In each section, 15\% of the data points were randomly selected to join the test dataset. The remaining 85\% of the data points form the training dataset. This strategy ensures that the statistical distribution of either the training or the test dataset is similar to that of the entire dataset. We designate this second approach of dividing the dataset as ``statistical splitting'', while the first approach is termed ``random splitting''.

\subsubsection{LASSO Regularization}

For a given polyimide, there were 1342 features generated for the dianhydride group and the same number of features for the diamine group. Not all these features play important roles in affecting the glass transition temperature of a polyimide. Including irrelevant or partially relevant features can lead to overfitting behavior of the resulting predictive model and negatively impact its performance. Overfitting is a common problem faced by machine-learning methods and many techniques have been developed to address this problem. In our approach, the importance of features were identified and ranked using the LASSO regularization method. At the end, a finite number of features were identified that control $T_\text{g}$ of polyimides.

In a linear fitting, each estimated target value $y_i$ could be represented as
\begin{equation}
\label{eq:linear_fit}
	y_i=\omega_0+\sum_{j=1}^px_{ij}\omega_j+\epsilon_i~,
\end{equation}
where $\omega_0$ is a constant, $\omega_j$ is a fitting parameter representing the coefficient of the $j$-th feature ($x_{ij}$) in a linear mapping from features to target value, $p$ is the number of features, and $\epsilon_i$ is the error of predicting the $i$-th data point. In a regular linear fitting scheme, the parameters $\omega_j$ can be found by minimizing the error function
\begin{equation}
	\text{error}=\sum_{i=1}^n(y_i-\omega_0-\sum_{j=1}^px_{ij}\omega_j)^2~,
\end{equation}
where $n$ is the number of data points. In the LASSO regularization method, the error to be minimized is slightly modified as
\begin{equation}
\label{en:lasso_error}
	\text{error}=\sum_{i=1}^n(y_i-\omega_0-\sum_{j=1}^px_{ij}\omega_j)^2+\lambda\sum_j^p|\omega_j|~,
\end{equation}
where $\lambda$ called a penalty factor. The advantage of the LASSO regularization method is that the coefficient of irrelevant and low-importance features can be shrunk to zero, which is an effective way of removing those features. If $\lambda$ is 0, then there will no shrinkage of any of the 1342 features, and LASSO regularization becomes linear regression.\cite{tibshirani1996regression} A big positive value of $\lambda$ indicates that the majority of the features will be removed. In the LASSO regularization method, $\lambda$ is therefore called a hyper-parameter which cannot be learned directly. In our implementation, the value of $\lambda$ was exhaustively searched from 0.01 to 2.0 in increments of 0.04. Our results reveal that the typical value of $\lambda$ is between 0.3 and 1.6. For such $\lambda$, most features are removed after LASSO regularization.  At the end, 197 features with nonzero coefficients remain in the final predictive model of $T_\text{g}$ (see the Supporting Information for the explanation of these 197 features) and the features with zero coefficients are removed during LASSO regularization. Out of 197 features, only about 12 features are actually important as indicated by their relatively large coefficients. The summation of the absolute value of coefficients of the largest 12 features is larger than the summation of the rest 185 features. Many of them can be easily justified on the basis of the available experimental evidence.

\subsubsection{Bagging}\label{sec:ch3_bagging}

Although the dataset on $T_\text{g}$ of polyimides has more data points than those in many previous studies on other classes of polymers,\cite{yu2006prediction, chen2008neural, ning2009artificial, xu2012prediction, pei2013modeling}, it is still a small set in the perspective of machine learning. The performance of the predictive model can exhibit significant fluctuations depending on how the dataset is split into a training and a test set. To reduce such variations, we utilized a bagging approach in the learning process.\cite{breiman1996bagging}

In the bagging approach, a dataset is randomly split into a training set and a test set. The machine-learning procedure described above, including the LASSO regularization method and an optimization process, is followed to generate a predictive model of $T_\text{g}$ using the training set. The whole process is then repeated by splitting the dataset into a new training set and a new test set. After $N_m$ repetitions, $N_m$ models are generated. The performance of each model is quantified by its error defined as
\begin{equation}
\label{eq:error}
	\text{Err}(k)=\sqrt{\frac{1}{n_k}\sum_{i=1}^{n_k}\left( T^k_\text{g}(i)_{\text{predicted}} -T^k_\text{g}(i)_{\text{target}}\right)^2}~,
\end{equation}
where $k$ is the index of the model and $n_k$ is the number of data points in the test set for the $k$-th model.

With the error associated with each model calculated, a weight, $\text{W}(k)$, was assigned to the $k$-th model according to
\begin{equation}
\label{eq:weight}
	\text{W}(k)=\cfrac{\left(\text{Err}(k)\right)^{-1}}{\sum_{j=1}^{N_m} \left(\text{Err}(j)\right)^{-1}}.
\end{equation}
The choice of the weight function in Eq.~(\ref{eq:weight}) guarantees that a model with a better performance, i.e., a smaller error in predicting the data in the corresponding test set, has a larger weight in the final predictive model. The final predictive model of $T_\text{g}$ is the linear combination of $N_m$ models weighted by $\text{W}(k)$ as in
\begin{equation}
\label{eq:final_predict}
         T_{\text{g}}=\sum_{k=1}^{N_m}\text{W}(k)\cdot T^k_{\text{g}}~.
\end{equation}
In the context of machine learning, this bagging procedure is often used to improve the robustness and stability of a learned model.

\subsection{Model Training and Test}

\subsubsection{Various Ways of Training Predictive Model of Glass Transition Temperature}

We implemented the machine-learning approach and tested the resulting predictive model of $T_\text{g}$ in four different ways. As discussed earlier, the dataset includes 151 data points from Ref.~\cite{ding2007isomeric}. In the first and second way, this dataset was randomly split into a test set containing 15\% of the data points. The remaining 85\% of the data formed the training set. In the first way, the training set was used to train the predictive model of $T_\text{g}$ via the LASSO regularization method but bagging was not used. In the LASSO regularization, the fitting parameter was exhaustively searched using a grid search method. The performance of the predictive model was quantified using the error of predicting the data points in the test set that never entered the training process. The entire procedure was repeated 1000 times and therefore 1000 models were generated. We analyzed the distribution of the errors of these models to predict the test set, which provided a metric quantifying the stability and performance of the first way of training the predictive model of $T_\text{g}$.

In the second way, ``random splitting'' was still used as in the first way but the bagging approach was used to train the predictive model of $T_\text{g}$ with each training set. In bagging, a training set was randomly split further into a training subset (85\%) and a test subset (15\%). The LASSO regularization method was applied to the training subset to obtain a model. This model was used to predict the test subset and the error of prediction was used to decide the weight (i.e., performance) of the model. The random splitting of the training set into two subsets was repeated 40 times, i.e., $N_m = 40$. The linear combination of these models yielded a blended predictive model of $T_\text{g}$. This blended model was used to predict the test set that never entered the training process and the associated error of prediction was taken as the gauge of the model's performance. The entire procedure was repeated 1000 times to generate 1000 blended predictive models.

The third and fourth ways were similar to the first and second ones except that ``statistical splitting'' discussed in Sec.~\ref{subsubsec:ch3_splitting} was used instead of ``random splitting''. In the third way, bagging was not used while the fourth way was a combination of ``statistical splitting'' and bagging. In all these ways, the first level of splitting was repeated 1000 times, resulting in 1000 models. When bagging was used, the second-level splitting of the initial training set into a training and a test subset was always repeated 40 times and therefore, all training approaches discussed here had $N_m = 40$.

\subsubsection{Performance of Predictive Model of Glass Transition Temperature}

\begin{figure}[htb]
    \centering
    \includegraphics[width=0.5\textwidth]{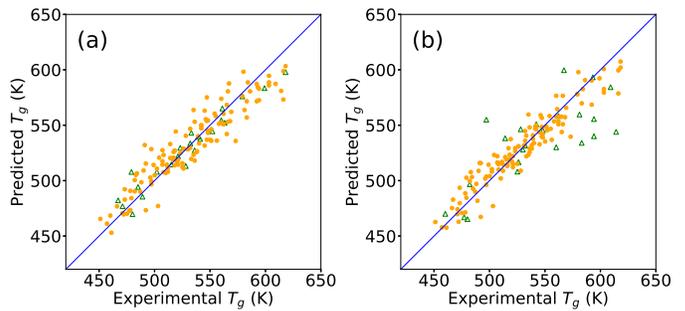}
    \caption{Performance of the (a) best and (b) worst model from training method \#1 (``random splitting'' + no bagging).}
    \label{fig:randomNobag}
\end{figure}

\begin{figure}[htb]
    \centering
    \includegraphics[width=0.5\textwidth]{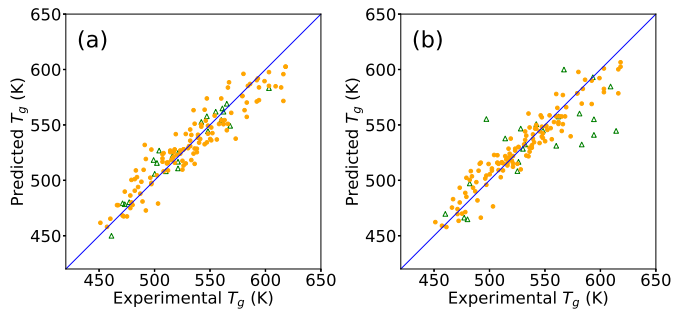}
    \caption{Performance of the (a) best and (b) worst model from training method \#2 (``random splitting'' + bagging).}
    \label{fig:randombag}
\end{figure}

\begin{figure}[htb]
    \centering
    \includegraphics[width=0.5\textwidth]{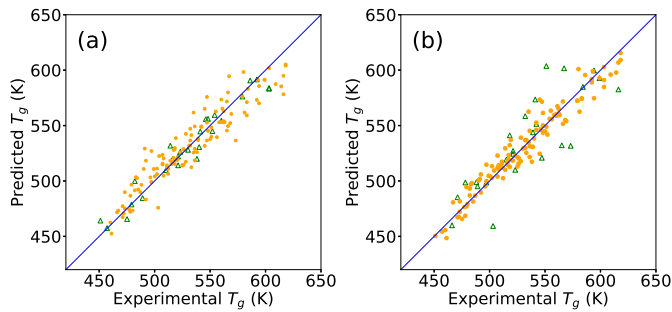}
    \caption{Performance of the (a) best and (b) worst model from training method \#3 (``statistical splitting'' + no bagging).}
    \label{fig:stdNobag}
\end{figure}

\begin{figure}[htb]
    \centering
    \includegraphics[width=0.5\textwidth]{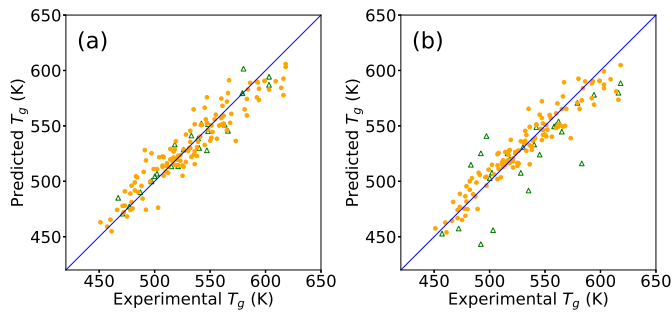}
    \caption{Performance of the (a) best and (b) worst model from training method \#4 (``statistical splitting'' + bagging).}
    \label{fig:stdbag}
\end{figure}

In this section, we show the performance of the predictive models of $T_\text{g}$ generated using the four training methods described previously. In the plots shown below, each data point represents one polyimide. For each point, the $x$-coordinate indicates the target, which is the actual value of $T_\text{g}$ determined experimentally. The $y$-coordinate indicates the predicted value of $T_\text{g}$ from a machine-learning-based model. The blue line indicates $y=x$. The closer a data point to the blue line, the better the performance of the predictive model. In Figs.~\ref{fig:randomNobag} to \ref{fig:stdbag}, orange dots represent the data used in training the predictive model while the green triangles represent the data used in testing the model, which did not enter the training process.

In our study, 1000 predictive models were generated using each training method. These models were ranked by their errors of predicting the test datasets that were not used to the model-training process. The performance of the best and worst predictive model of $T_\text{g}$ derived in the first manner of implementing the machine-learning approach described previously (i.e., ``random splitting'' + no bagging) is shown in Fig.~\ref{fig:randomNobag}. The error of using the best model to predict the training set is 14.37 K while the error is 10.78 K if the model is used to predict the test dataset. The errors of the worst model are 12.03 K for the training set and 29.62 K for the test set, respectively. The large prediction error for the test set indicates that the corresponding model has a poor prediction power.

The performance of the best and worst predictive model of $T_\text{g}$ trained with the second method (i.e., ``random splitting'' + bagging) is shown in Fig.~\ref{fig:randombag}. The error of predicting the training set is 14.26 K for the best model and is 12.08 K for the worst model. The best model has an error of 10.65 K when it is used to predict the test dataset while the prediction error is much larger at 29.61 K when the worst model is used.

Figure \ref{fig:stdNobag} shows the performance of the best and worst models trained using the third method (i.e., ``statistical splitting'' + no bagging). The best model has an error of 13.65 K of predicting the training set and of 9.79 K for the test dataset. The worst model has a smaller error at 9.66 K of predicting the training set, which indicates that the model-training is successful. However, the error is much larger at 30.05 K when the test dataset was used to check the performance of the predictive model. This large discrepancy of errors of predicting the training and test dataset is a reflection of the overfitting issue faced by many machine-learning approaches. Below we show that bagging can be used to effectively address this issue.

The bagging approach introduced in Sec.~\ref{sec:ch3_bagging} can be used to improve the stability of a machine-learning-trained predictive model. In Fig.~\ref{fig:stdbag}, we show the performance of the best and worst model trained with the fourth method that combines ``statistical splitting'' of the dataset with a bagging approach. In Fig.~\ref{fig:stdbag}(a), the errors of the best model of predicting the training and test dataset are 13.80 K and 10.21 K, respectively. For the worst model, the corresponding errors are 11.86 K and 27.37 K, as shown in Fig.~\ref{fig:stdbag}(b).

To quantitatively compare the various ways of training the predictive model of $T_\text{g}$, we performed a statistical analysis of the errors of the 1000 models when they were used to predict the test dataset. The average and standard deviation of these errors are included in Table~\ref{tb:model_performance}. The results show that when bagging is used, both average and standard deviation of the errors are reduced. Bagging thus enhances the stability of the machine-learning-trained model. Furthermore, the training methods in which ``statistical splitting'' is used to make the training dataset more statistically representative of the entire dataset also lead to predictive models with better performance. The trends indicate that the best training method is to use ``statistical splitting'' coupled with bagging, i.e, the fourth method.

\begin{table*}[htb]
\centering
\caption{Performance metrics of the predictive models of $T_\text{g}$}
\label{tb:model_performance}
\begin{tabular}{|c|c|c|c|}
\hline
Training Method & Average Error (K) & Standard Deviation of Error (K) & Correlation \\ \hline
Method \#1 & 18.58 & 3.09 & 0.71 \\ \hline
Method \#2 & 18.31 & 2.98 & 0.84 \\ \hline
Method \#3 & 18.17 & 2.87 & 0.68 \\ \hline
Method \#4 & 17.98 & 2.62 & 0.83 \\ \hline
\end{tabular}
\end{table*}

In each splitting of the entire dataset into a training and a test set, a predictive model of $T_\text{g}$ was generated. This model is a linear mapping from all $Z$ features generated for the dianhydride and diamine groups to $T_\text{g}$, with the coefficient of $k$-th feature denoted as $M(k)$. A larger absolute value of $M(k)$ implies that the corresponding $k$-th feature is more strongly correlated to $T_\text{g}$. For two models, a correlation can thus be defined as
\begin{equation}\label{eq:corrmodel}
    c_{ij} = \frac{2\sum^Z_{k=1}M_i(k)M_j(k)}{\sum^Z_{k=1}\left[ M_i^2(k)+M_j^2(k) \right]}~,
\end{equation}
where $i$ and $j$ are the indices of the models. If the two models are identical, then $c_{ij} = 1$. If the two models are anticorrelated with $M_i(k) = -M_j(k)$, then $c_{ij} = -1$. If a training method of the predictive model of $T_\text{g}$ is stable, then different splittings will lead to models that are highly correlated, with the correlation between the models close to 1.

We computed the correlations of all pairs out of the 1000 models generated with one of the four training methods discussed previously, using Eq.~(\ref{eq:corrmodel}). The average correlation for each training method is included in Table~\ref{tb:model_performance}. It is clear that the bagging method significantly increases the correlation between the resulting predictive models and thus enhances the stability of the training process.

\begin{figure}
    \centering
    \includegraphics[width=0.45\textwidth]{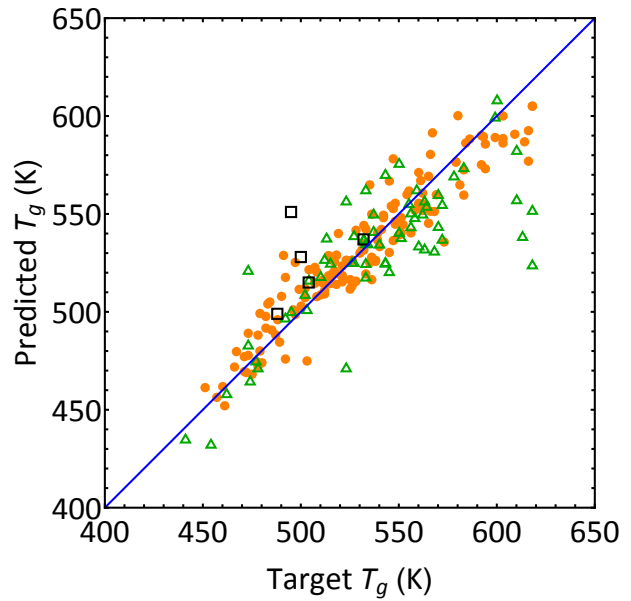}
    \caption[cap for figure]{Performance of the predictive model of $T_\text{g}$ trained with 151 data points (orange dots) from Ref.~\cite{ding2007isomeric} and tested against 63 data points (green triangles) from Ref.~\cite{liu2010prediction} as well as 5 polyetherimides (black squares) in Table~\ref{tb:TgData}. For these 5 polyetherimides, the predicted $T_\text{g}$ is plotted against the target value determined with $D(T)$ from all-atom MD simulations. For data from Ref.~\cite{liu2010prediction}, the target $T_\text{g}$ represents the experimental value.}
    \label{fig:indep}
\end{figure}

Finally, we used the fourth method (i.e., ``statistical splitting'' + bagging) to train a predictive model of $T_\text{g}$ with all 151 data points from Ref.~\cite{ding2007isomeric}. The model was then used to predict the 63 data points from Ref.~\cite{liu2010prediction}. Since the training and test datasets in this case are from two different sources, this test serves as an independence check of the training method. The resulting predictive model of $T_\text{g}$ has an average error of 25.5 K of predicting the test set, as shown in Fig.~\ref{fig:indep}. However, large errors mainly occur for high $T_\text{g}$ around 615 K. In the lower range of $T_\text{g}$, Fig.~\ref{fig:indep} indicates that the predictive model performs well in terms of predicting the independent dataset from a different source. The coefficients of all 197 features that enter the predictive model are included in the Supporting Information. Among them 12 features are found to be the most important ones. The details of these feature are also available in the Supporting Information.

We further applied the predictive model to predict $T_\text{g}$ for the 5 polyetherimides in Table~\ref{tb:TgData}. Out of this group, 4,4'BPADA+DDS, 4,4'BPADA+CDA, and 3,3'BPADA+MPD were made and characterized after the prediction. For 4,4'BPADA+PPD, the experimental values of $T_\text{g}$ is still unavailable as it is not synthesized yet. Therefore, we plot the predicted $T_\text{g}$ against the value determined with $D(T)$ from all-atom MD simulations in Fig.~\ref{fig:indep}. The results show that except for 4,4'BPADA+CDA, the predictive model yields estimates close to the target values. The prediction errors are about 10 to 20 K, which are comparable to those of using the same model to predict the training dataset and even smaller than those of the independent test dataset from a different source. This comparison further validates the prediction power of the model constructed via the machine-learning approach. It also shows that a $D(T)$ curve computed with atomistic MD simulations can be used to estimate $T_\text{g}$ with reasonable accuracy.

\section{Conclusions}\label{sec:ch3_conclusion}

In this paper we show that the PCFF force field combined with Mulliken charges can be used in all-atom MD simulations to compute and estimate $T_\text{g}$ of polyimides. The determination of $T_\text{g}$ can be achieved by computing either the polymer density, $\rho$, or diffusion coefficients of gas molecules, $D$, in the polymer matrix as a function of temperature. For temperatures lower or higher than $T_\text{g}$, $\rho$ exhibits a linear dependence on $T$ but the slopes are different. $D$, on the other hand, depends on $T$ as $\log D \propto 1/T$ and the linear coefficients are again different for $T < T_\text{g}$ and $T > T_\text{g}$. The comparison shows that in practice, $D$ can be more reliably computed and used to give a more accurate estimate of $T_\text{g}$. However, several limitations of using all-atom MD simulations to compute $T_\text{g}$ should be noted. First, the cooling rate used in MD simulations is typically several orders of magnitude larger than experimental rates. Secondly, the molecular weight of the polymers modeled in all-atom MD simulations is usually smaller than experimental values by a factor of 10 to 100. Thirdly, it is challenging to study a polydisperse system in MD simulations. Lastly, the PCFF force field is a generic force field for polymers and not specifically designed and optimized for polyimides. All these issues point to the need of going beyond all-atom MD simulations and seeking a predictive model that can be used to quickly estimate $T_\text{g}$ of polyimides.

A predictive model of $T_\text{g}$ of polyimides can be obtained by applying machine-learning algorithms to analyze available experimental and simulation data on $T_\text{g}$. We demonstrate a machine-learning approach to systematically derive such predictive models, including using a SMILES notation to designate a polymer, feature generation by reading in the SMILES notation, removal of irrelevant and low-importance features through the LASSO regularization method, and improving and optimizing the predictive models via bagging. For polyimides, we have explored 4 different training methods to construct a predictive model of $T_\text{g}$ of polyimides using data collected from Ref.~\cite{ding2007isomeric} and found that the best model is obtained if the entire dataset is split into the training and test sets that are statistically representative of the entire set and if bagging is used to improve the stability of the predictive model. We further demonstrate that this model can be successfully applied to accurately predict the results on $T_\text{g}$ reported in Ref.~\cite{liu2010prediction}, which were from a different source and never used to train the predictive model. Furthermore, even for polyimides that are yet to be synthesized, the predictive model yields value of $T_\text{g}$ close to those determined with all-atom MD simulations, which validates the prediction power of the model. In the future, it is interesting to further improve the predictive model of $T_\text{g}$ by training it with a larger dataset and taking into account the differences in the experimental conditions under which $T_\text{g}$ is measured. It is also interesting to explore if similar predictive models of other physical quantities of interest, such as dielectric constants and mechanical moduli, can be developed for polyimides.

\section*{Acknowledgments}
This paper is based on the results from work supported by SABIC Innovative Plastics US LLC. The authors acknowledge Advanced Research Computing at Virginia Tech (URL: http://www.arc.vt.edu) for providing computational resources and technical support that have contributed to the results reported within this paper. The authors also gratefully acknowledge the support of NVIDIA Corporation with the donation of the Tesla K40 GPU used for this research.


\onecolumngrid



\section*{Supplementary material}

# Supporting Information:

# "Determination of Glass Transition Temperature of Polyimides from Atomistic Molecular Dynamics Simulations and Machine-Learning Algorithms"

Chengyuan Wen,[1,2] Binghan Liu,[1,2] Josh Wolfgang,[2,3]

Timothy E. Long,[2,3] Roy Odle,[4] and Shengfeng Cheng[1,2,5,*]

[1]*Department of Physics and Center for Soft Matter and Biological Physics,*

*Virginia Polytechnic Institute and State University, Blacksburg, Virginia 24061, USA*

[2]*Macromolecules Innovation Institute, Virginia Polytechnic*

*Institute and State University, Blacksburg, Virginia 24061, USA*

[3]*Department of Chemistry, Virginia Polytechnic Institute*

*and State University, Blacksburg, Virginia 24061, USA*

[4]*SABIC Innovative Plastics, 1 Lexan Lane, Mt. Vernon, Indiana 47620, USA*

[5]*Department of Mechanical Engineering,*

*Virginia Polytechnic Institute and State University, Blacksburg, Virginia 24061, USA*

## S1. Additional Results on Polymer Density and Gas Diffusion Coefficient

Figures S1 to S8 show the results from all-atom molecular dynamics simulations on density ($\rho$) of polymer and diffusion coefficient ($D$) of gas molecules (either argon or neon) as a function of temperature for the remaining 4 polyetherimides included in Table 1 of the main text. These results are used to determine their respective glass transition temperature, $T_g$, shown in Table 1 of the main text.

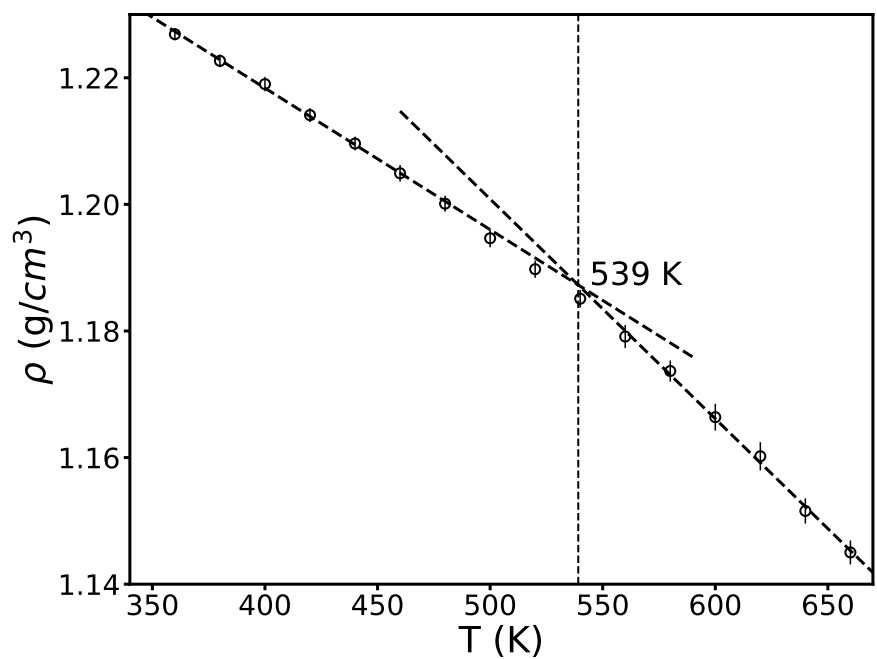

FIG. S1. $\rho(T)$ vs. $T$ for 4,4'BPADA+PPD, for which $T_g = 539$ K.

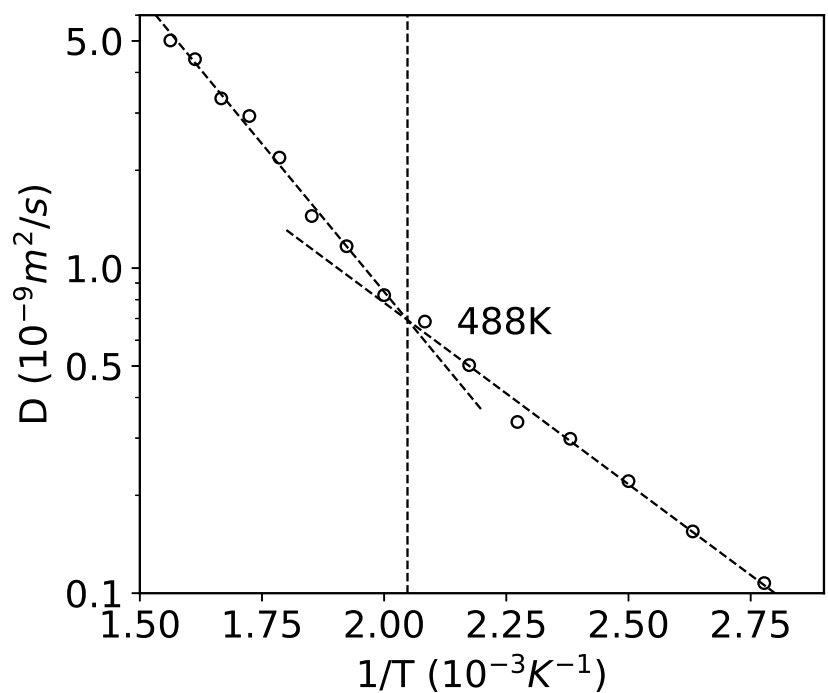

FIG. S2. $D(T)$ vs. $1/T$ on a log-linear scale for argon diffusing in 4,4'BPADA+PPD, which yields $T_g = 488$ K.

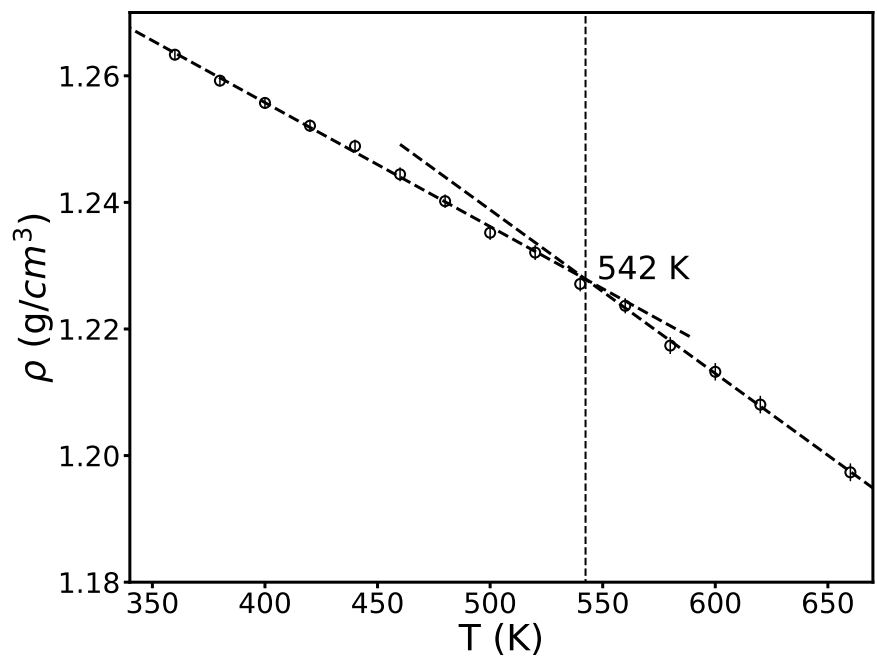

FIG. S3. $\rho(T)$ vs. $T$ for 4,4'BPADA+DDS, for which $T_g = 542$ K.

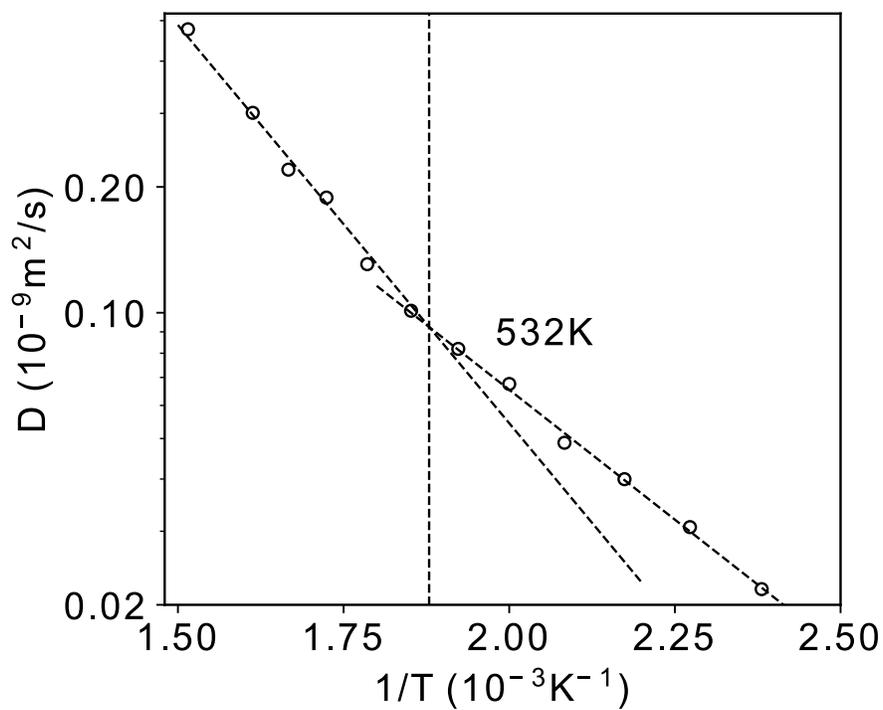

FIG. S4. $D(T)$ vs. $1/T$ on a log-linear scale for neon diffusing in 4,4'BPADA+DDS, which yields $T_g = 532$ K.

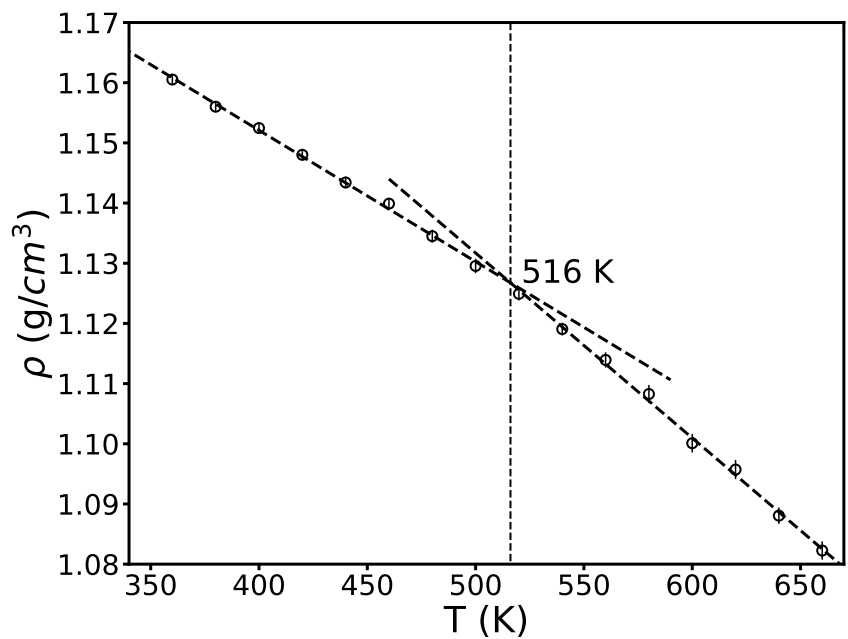

FIG. S5. $\rho(T)$ vs. $T$ for 4,4'BPADA+CDA, for which $T_\mathrm{g} = 516$ K.

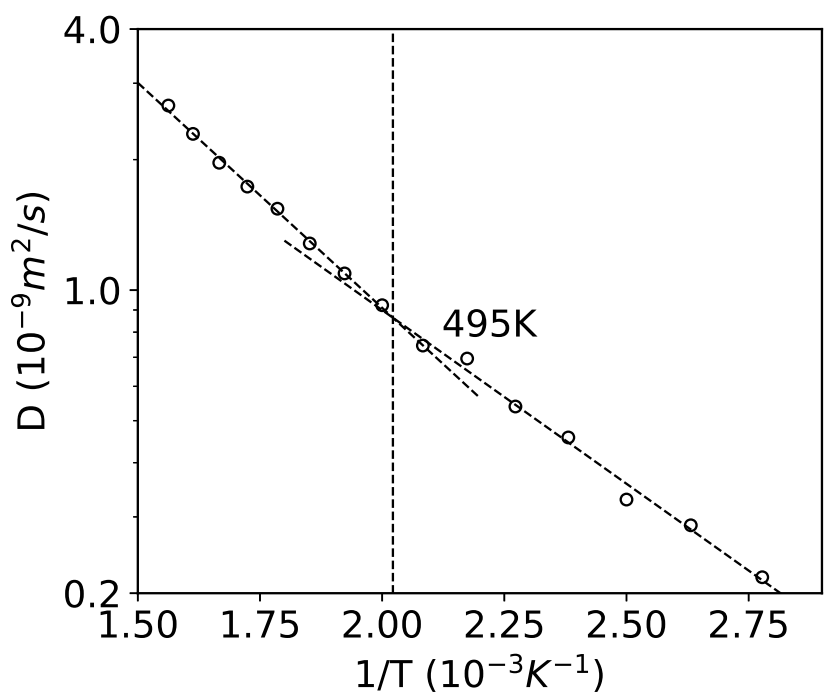

FIG. S6. $D(T)$ vs. $1/T$ on a log-linear scale for neon diffusing in 4,4'BPADA+CDA, which yields $T_\mathrm{g} = 495$ K.

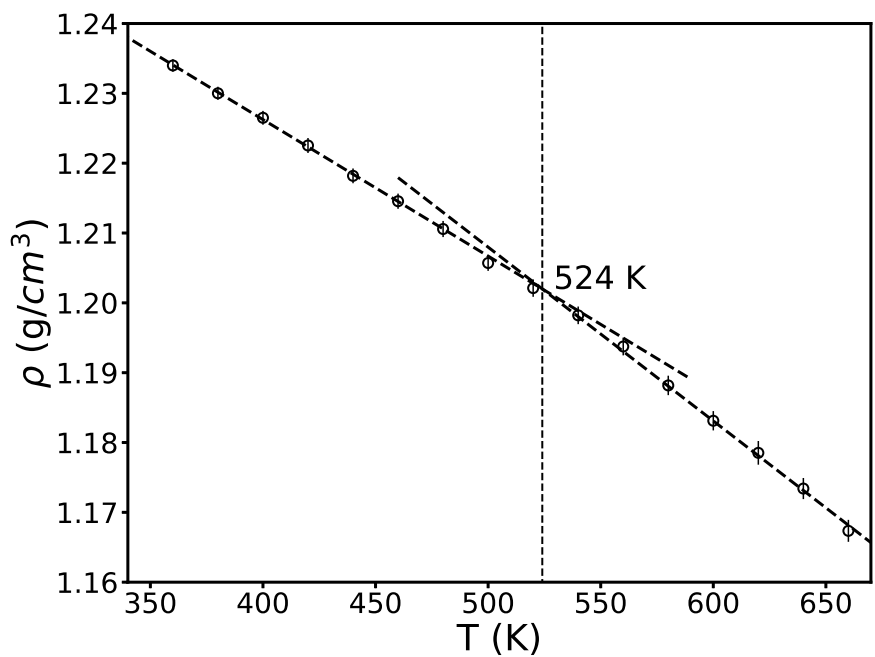

FIG. S7. $\rho(T)$ vs. $T$ for 3,3'BPADA+MPD, for which $T_g = 524$ K.

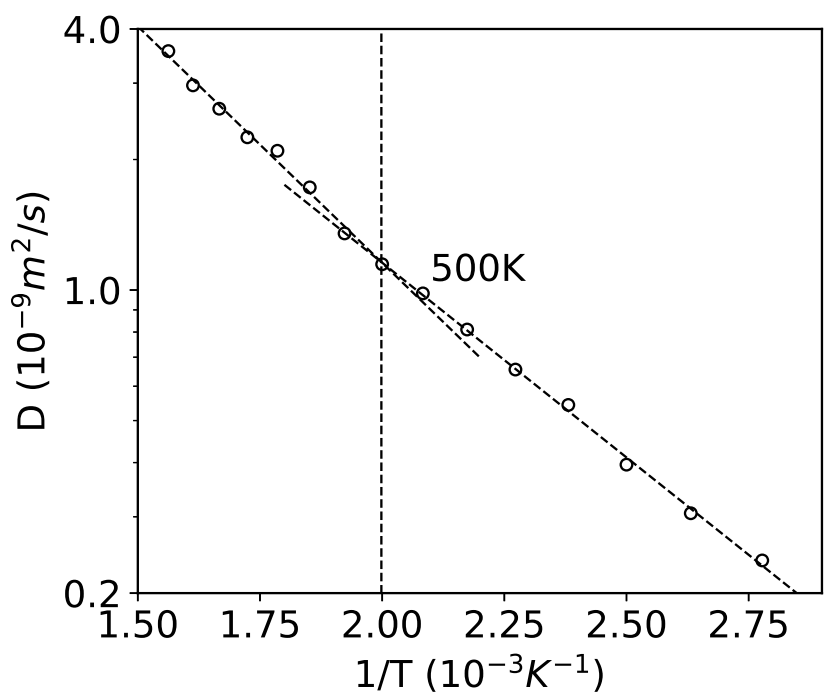

FIG. S8. $D(T)$ vs. $1/T$ on a log-linear scale for neon diffusing in 3,3'BPADA+MPD, which yields $T_g = 500$ K.

**S2: Definition of Polymer Features**

The definition of 197 features (i.e., descriptors) of polyimides, which enter the final predictive model of $T_{\mathrm{g}}$, are shown in Table S1.

TABLE S1: Definition of features.

| Constitutional descriptors | |
|---|---|
| Constitutional descriptors are OD-descriptors, independent from molecular connectivity and conformations, Atom and bond counts, molecular weight sum of atomic properties, ets. | |
| nDB | number of double bonds |
| PJI2 | 2D Petitjean shape index |
| | |
| Topological descriptors | |
| Topological descriptors are molecular descriptors obtained from molecular graph (usually H-depleted), i.e. 2D-descriptors conformationally independent. | |
| TNN | sum of topological distances between N..N |
| TOO | sum of topological distances between O..O |
| TOS | sum of topological distances between O..S |
| | |
| Walk and path counts | |
| Walk and path counts are molecular descriptors obtained from the molecular graph, counting paths, walks and self-returning walks of different lengths. | |
| SRW09 | self-returning walk count of order 9 |
| | |
| Information indices | |
| Information indices are molecular descriptors calculated as information content of molecules, based on the calculation of equivalence classes from the molecular graph. Among them, the indices of neighborhood symmetry take into account also neighbor degree and edge multiplicity. | |

| | |
|---|---|
| IC2 | Information Content index (neighborhood symmetry of 3-order) |
| IC3 | Information Content index (neighborhood symmetry of 3-order) |
| BIC3 | Bond Information Content index (neighborhood symmetry of 3-order) |
| | |
| 2D autocorrelations | |
| Molecular descriptors calculated from molecular graph by summing the products of atom weights of the terminal atoms of all the paths of the considered path length (the lag). 2D autocorrelations by Moreau-Broto (ATS), Moran (MATS) and Geary (GATS) algorithms are calculated from lag 1 to lag 8 for 4 different weighting schemes. | |
| MATS1m | Moran autocorrelation of lag 1 weighted by mass |
| MATS4m | Moran autocorrelation of lag 4 weighted by mass |
| MATS8m | Moran autocorrelation of lag 8 weighted by mass |
| MATS4v | Moran autocorrelation of lag 4 weighted by van der Waals volume |
| MATS6v | Moran autocorrelation of lag 6 weighted by van der Waals volume |
| MATS8v | Moran autocorrelation of lag 8 weighted by van der Waals volume |
| MATS4e | Moran autocorrelation of lag 4 weighted by Sanderson electronegativity |
| MATS4p | Moran autocorrelation of lag 4 weighted by polarizability |
| MATS6p | Moran autocorrelation of lag 6 weighted by polarizability |

| | |
|---|---|
| MATS7p | Moran autocorrelation of lag 7 weighted by polarizability |
| GATS4m | Geary autocorrelation of lag 4 weighted by mass |
| GATS5m | Geary autocorrelation of lag 5 weighted by mass |
| GATS8m | Geary autocorrelation of lag 8 weighted by mass |
| GATS3e | Geary autocorrelation of lag 3 weighted by Sanderson electronegativity |
| GATS4e | Geary autocorrelation of lag 4 weighted by Sanderson electronegativity |
| GATS8e | Geary autocorrelation of lag 8 weighted by Sanderson electronegativity |
| GATS4p | Geary autocorrelation of lag 4 weighted by polarizability |
| GATS5p | Geary autocorrelation of lag 5 weighted by polarizability |
| | |
| BCUT descriptors | |
| BCUT descriptors are molecular descriptors obtained from the positive and negative eigenvalues of the adjacency matrix, weighting the diagonal elements with atom weights. | |
| BELm4 | lowest eigenvalue n. 4 of Burden matrix / weighted by atomic masses |
| BELv5 | lowest eigenvalue n. 5 of Burden matrix / weighted by atomic van der Waals volumes |

| | |
|---|---|
| BELp5 | lowest eigenvalue n. 5 of Burden matrix / weighted by atomic polarizabilities |
| | |
| Topological charge indices | |
| First 10 eigenvalues (absolute values) obtained from a corrected adjacency matrix, i.e. diagonal elements correspond to .... | |
| GGI8 | topological charge index of order 8 |
| GGI9 | topological charge index of order 9 |
| GGI10 | topological charge index of order 10 |
| JGI1 | mean topological charge index of order 1 |
| JGI2 | mean topological charge index of order 2 |
| JGI3 | mean topological charge index of order 3 |
| JGI4 | mean topological charge index of order 4 |
| JGI7 | mean topological charge index of order 7 |
| JGI8 | mean topological charge index of order 8 |
| JGI9 | mean topological charge index of order 9 |
| JGI10 | mean topological charge index of order 10 |
| | |
| Eigenvalue-based indices | |
| Topological descriptors calculated by the eigenvalues of a square (usually symmetric) matrix representing a molecular graph. | |
| VEA1 | eigenvector coefficient sum from adjacency matrix |
| | |
| Geometrical descriptors | |

| | |
|---|---|
| Geometrical descriptors are different kinds of conformationally dependent descriptors based on the molecular geometry. Reliable values are obtained if reliable conformations were previously calculated. | |
| SPAM | average span R |
| MEcc | molecular eccentricity |
| FDI | folding degree index |
| PJI3 | 3D Petitjean shape index |
| LBw | length-to-breadth ratio by WHIM |
| HOMA | Harmonic Oscillator Model of Aromaticity index |
| RCI | 3D Petitjean shape index |
| AROM | aromaticity index |
| DISPv | displacement value / weighted by van der Waals volume |
| GNN | sum of geometrical distances between N..N |
| GNF | sum of geometrical distances between N..F |
| GOS | sum of geometrical distances between O..S |
| | |
| RDF Descriptors | |
| RDF Descriptors are Molecular descriptors obtained by radial basis functions centered on different interatomic distances (from 0.5A to 15.5A). | |
| RDF110u | Radial Distribution Function - 110 / unweighted |
| RDF135u | Radial Distribution Function - 135 / unweighted |
| RDF140u | Radial Distribution Function - 140 / unweighted |
| RDF145u | Radial Distribution Function - 145 / unweighted |
| RDF045m | Radial Distribution Function - 045 / weighted by mass |
| RDF055m | Radial Distribution Function - 055 / weighted by mass |
| RDF075m | Radial Distribution Function - 075 / weighted by mass |

| RDF080m | Radial Distribution Function - 080 / weighted by mass |
|---------|-------------------------------------------------------|
| RDF085m | Radial Distribution Function - 085 / weighted by mass |
| RDF090m | Radial Distribution Function - 090 / weighted by mass |
| RDF095m | Radial Distribution Function - 095 / weighted by mass |
| RDF100m | Radial Distribution Function - 100 / weighted by mass |
| RDF105m | Radial Distribution Function - 105 / weighted by mass |
| RDF125m | Radial Distribution Function - 125 / weighted by mass |
| RDF130m | Radial Distribution Function - 130 / weighted by mass |
| RDF135m | Radial Distribution Function - 135 / weighted by mass |
| RDF145m | Radial Distribution Function - 145 / weighted by mass |
| RDF150m | Radial Distribution Function - 150 / weighted by mass |
| RDF065v | Radial Distribution Function - 065 / weighted by van der Waals volume |
| RDF135v | Radial Distribution Function - 135 / weighted by van der Waals volume |
| RDF150v | Radial Distribution Function - 150 / weighted by van der Waals volume |
| RDF155v | Radial Distribution Function - 155 / weighted by van der Waals volume |
| RDF090e | Radial Distribution Function - 090 / weighted by Sanderson electronegativity |
| RDF120e | Radial Distribution Function - 120 / weighted by Sanderson electronegativity |
| RDF145e | Radial Distribution Function - 145 / weighted by Sanderson electronegativity |
| RDF135p | Radial Distribution Function - 135 / weighted by polarizability |
| RDF150p | Radial Distribution Function - 150 / weighted by polarizability |

| RDF155p | Radial Distribution Function - 155 / weighted by polarizability |
|---------|------------------------------------------------------------------|
|         |                                                                  |

| 3D-MoRSE descriptors |
|----------------------|

| 3D-MoRSE is a very flexible 3D structure encoding framework for chemoinformatics and QSAR purposes due to the range of scattering parameter values and variety of weighting schemes used. |
|---|

| Mor02u | signal 02 / unweighted |
|--------|------------------------|
| Mor04u | signal 04 / unweighted |
| Mor05u | signal 05 / unweighted |
| Mor08u | signal 08 / unweighted |
| Mor09u | signal 09 / unweighted |
| Mor12u | signal 12 / unweighted |
| Mor13u | signal 13 / unweighted |
| Mor16u | signal 16 / unweighted |
| Mor17u | signal 17 / unweighted |
| Mor21u | signal 21 / unweighted |
| Mor22u | signal 22 / unweighted |
| Mor24u | signal 24 / unweighted |
| Mor25u | signal 25 / unweighted |
| Mor28u | signal 28 / unweighted |
| Mor03m | signal 03 / weighted by mass |
| Mor11m | signal 11 / weighted by mass |
| Mor16m | signal 16 / weighted by mass |
| Mor23m | signal 23 / weighted by mass |
| Mor25m | signal 25 / weighted by mass |
| Mor26m | signal 26 / weighted by mass |
| Mor28m | signal 28 / weighted by mass |

| Mor29m | signal 29 / weighted by mass |
|--------|------------------------------|
| Mor03v | signal 03 / weighted by van der Waals volume |
| Mor06v | signal 06 / weighted by van der Waals volume |
| Mor17v | signal 17 / weighted by van der Waals volume |
| Mor21v | signal 21 / weighted by van der Waals volume |
| Mor22v | signal 22 / weighted by van der Waals volume |
| Mor24v | signal 24 / weighted by van der Waals volume |
| Mor28v | signal 28 / weighted by van der Waals volume |
| Mor29v | signal 29 / weighted by van der Waals volume |
| Mor30v | signal 30 / weighted by van der Waals volume |
| Mor02e | signal 02 / weighted by van der Waals volume |
| Mor03e | signal 03 / weighted by van der Waals volume |
| Mor08e | signal 08 / weighted by van der Waals volume |
| Mor11e | signal 11 / weighted by van der Waals volume |
| Mor14e | signal 14 / weighted by van der Waals volume |
| Mor15e | signal 15 / weighted by van der Waals volume |
| Mor17e | signal 17 / weighted by van der Waals volume |
| Mor18e | signal 18 / weighted by van der Waals volume |
| Mor21e | signal 21 / weighted by van der Waals volume |
| Mor25e | signal 25 / weighted by van der Waals volume |
| Mor30e | signal 30 / weighted by van der Waals volume |
| Mor03p | signal 03 / weighted by polarizability |
| Mor04p | signal 04 / weighted by polarizability |
| Mor10p | signal 10 / weighted by polarizability |
| Mor17p | signal 17 / weighted by polarizability |
| Mor24p | signal 24 / weighted by polarizability |
| Mor29p | signal 29 / weighted by polarizability |
|        |  |
| WHIM descriptors |  |

| | WHIM descriptors are molecular descriptors obtained as statistical indices of the atoms projected onto the 3 principal components obtained from weighted covariance matrices of the atomic coordinates. | |
|---|---|---|
| G1u | 1st component symmetry directional WHIM index / unweighted | |
| G3u | 3rd component symmetry directional WHIM index / unweighted | |
| E1u | 1st component accessibility directional WHIM index / unweighted | |
| E2u | 2ed component accessibility directional WHIM index / unweighted | |
| E3u | 3rd component accessibility directional WHIM index / unweighted | |
| L3m | 3rd component size directional WHIM index / weighted by atomic masses | |
| G1m | 1st component symmetry directional WHIM index / weighted by atomic masses | |
| G2m | 2ed component symmetry directional WHIM index / weighted by atomic masses | |
| E1m | 1st component accessibility directional WHIM index / weighted by atomic masse | |
| E1v | 1st component accessibility directional WHIM index / weighted by atomic van der Waals volumes | |
| P2e | 2nd component shape directional WHIM index / weighted by atomic Sanderson electronegativities | |
| G2e | 2st component symmetry directional WHIM index / weighted by atomic Sanderson electronegativities | |
| E1e | 1st component accessibility directional WHIM index / weighted by atomic Sanderson electronegativities | |
| E2e | 2nd component accessibility directional WHIM index / weighted by atomic Sanderson electronegativities | |
| G2p | 2st component symmetry directional WHIM index / weighted by atomic polarizabilities | |

| G3p | 3st component symmetry directional WHIM index / weighted by atomic polarizabilities |
|---|---|
| E1p | 1st component accessibility directional WHIM index / weighted by atomic polarizabilities |
| E2p | 2ed component accessibility directional WHIM index / weighted by atomic polarizabilities |
| L3s | 3rd component size directional WHIM index / weighted by atomic electrotopological states |
| E1s | 1st component accessibility directional WHIM index / weighted by atomic electrotopological states |
| Gu | G total symmetry index / unweighted |
| Du | D total accessibility index / unweighted |
| Dm | D total accessibility index / weighted by atomic masses |
| Ds | D total accessibility index / weighted by atomic electrotopological states |
| | |
| GETAWAY descriptors | |
| GETAWAY descriptors are descriptors calculated from the leverage matrix obtained by the centered atomic coordinates (molecular influence matrix, MIM). The first four descriptors are calculated as information content and connectivity indices. HATS and H descriptors are 3D-autocorrelation descriptors obtained from MIM; R and R+ descriptors are analogously obtained from the leverage/geometry matrix. | |
| ISH | standardized information content on the leverage equality |
| H7u | H autocorrelation of lag 7 / unweighted |
| H8u | H autocorrelation of lag 8 / unweighted |
| HATS7u | leverage-weighted autocorrelation of lag 7 / unweighted |
| H8m | H autocorrelation of lag 8 / weighted by atomic masses |

| | |
|---|---|
| HATS8m | leverage-weighted autocorrelation of lag 8 / weighted by atomic masses |
| H8e | H autocorrelation of lag 8 / weighted by atomic Sanderson electronegativities |
| R2u+ | R maximal autocorrelation of lag 2 / unweighted |
| R6u+ | R maximal autocorrelation of lag 6 / unweighted |
| R7u+ | R maximal autocorrelation of lag 7 / unweighted |
| R5m+ | R maximal autocorrelation of lag 5 / weighted by atomic masses |
| R7m+ | R maximal autocorrelation of lag 7 / weighted by atomic masses |
| R8m+ | R maximal autocorrelation of lag 8 / weighted by atomic masses |
| R1v+ | R maximal autocorrelation of lag 1 / weighted by atomic van der Waals volumes |
| R2v+ | R maximal autocorrelation of lag 2 / weighted by atomic van der Waals volumes |
| R3v+ | R maximal autocorrelation of lag 3 / weighted by atomic van der Waals volumes |
| R6v+ | R maximal autocorrelation of lag 6 / weighted by atomic van der Waals volumes |
| R5e+ | R maximal autocorrelation of lag 5 / weighted by atomic Sanderson electronegativities |
| R7e+ | R maximal autocorrelation of lag 7 / weighted by atomic Sanderson electronegativities |
| R8p | R autocorrelation of lag 8 / weighted by atomic polarizabilities |
| R1p+ | R maximal autocorrelation of lag 1 / weighted by atomic polarizabilities |

| R2p+ | R maximal autocorrelation of lag 2 / weighted by atomic polarizabilities |
|------|---------------------------------------------------------------------------|
| R23+ | R maximal autocorrelation of lag 3 / weighted by atomic polarizabilities |
| R5p+ | R maximal autocorrelation of lag 5 / weighted by atomic polarizabilities |
| R6p+ | R maximal autocorrelation of lag 6 / weighted by atomic polarizabilities |
| R8p+ | R maximal autocorrelation of lag 8 / weighted by atomic polarizabilities |
| RTp+ | R maximal index / weighted by atomic polarizabilities |

|  |  |
|--|--|

| Functional group counts | |
|-------------------------|--|
| Molecular descriptors based on the counting of chemical functional groups. | |
| nCs | number of total secondary C(sp3) |
| nCconj | number of exo-conjugated C(sp2) |
| nArCO | number of ketones (aromatic) |
| nHBonds | number of intramolecular H-bonds (with N,O,F) |

|  |  |
|--|--|

| Atom-centred fragments | |
|------------------------|--|
| Atom-centred fragments are molecular descriptors based on the counting of 120 atom-cantered fragments, as defined by Ghose-Crippen. Some fragments are undefined by the authors. | |
| C-039 | |
| O-058 | |

|  |  |
|--|--|

| Molecular properties | |
|----------------------|--|
| Molecular properties calculated from models together with some empirical descriptors. | |

| LAI | Lipinski Alert index |
|-----|----------------------|
| GVWAI-50 | Ghose-Viswanadhan-Wendoloski Alert index at 50% (drug-like index) |
| Inflammat-50 | Ghose-Viswanadhan-Wendoloski antiinflammatory-like index at 50% |
| Depressant-80 | Ghose-Viswanadhan-Wendoloski antidepressant at 80% (drug-like index) |
| Psychotic-80 | Ghose-Viswanadhan-Wendoloski antipsychotic at 80% (drug-like index) |
| Psychotic-50 | Ghose-Viswanadhan-Wendoloski antipsychotic at 50% (drug-like index) |
| Neoplastic-50 | Ghose-Viswanadhan-Wendoloski antineoplastic at 50% (drug-like index) |

**S3: Predictive Model of Glass Transition Temperature of Polyimides from Machine Learning**

Table S2 shows 197 features and their coefficients in the predictive model of $T_g$ of polyimides trained with 151 data points from Ref. [1] using the training method based on "statistical splitting" and bagging discussed in the main text.

TABLE S2: Linear coefficients of features.

| Feature | Coefficient | Feature | Coefficient |
|---|---|---|---|
| Mor04u | 13.046288 | IC3 | -5.176466 |
| JGI2 | 9.477253 | MATS4m | -3.955315 |
| Mor11m | 3.543599 | VEA1 | -3.682337 |
| GVWAI-50 | 3.024074 | Mor26m | -3.398099 |
| Mor17p | 1.755194 | R1p+ | -3.056214 |
| Mor22v | 1.721365 | E1v | -2.842052 |
| Mor03v | 1.662579 | Mor02e | -2.532373 |
| MEcc | 1.508814 | R1v+ | -2.327931 |
| Mor17e | 1.23228 | RDF130m | -2.202375 |
| MATS6v | 1.117188 | PJI2 | -1.651294 |
| R7e+ | 1.037976 | JGI3 | -1.553203 |
| Mor17u | 1.027713 | GGI10 | -1.425599 |
| Neoplastic-50 | 0.974793 | Mor29m | -1.285398 |
| nHBonds | 0.782506 | Mor02u | -1.02832 |
| Mor24p | 0.760354 | Mor25e | -0.947649 |
| RDF120e | 0.749853 | ISH | -0.93868 |
| Mor03p | 0.729866 | MATS4e | -0.932648 |
| JGI9 | 0.709341 | RDF075m | -0.828565 |
| Mor17v | 0.675846 | MATS4v | -0.806 |
| Mor03m | 0.626338 | RDF100m | -0.764006 |
| R7u+ | 0.570961 | MATS4p | -0.638701 |
| E2u | 0.570157 | Mor25u | -0.63294 |
| nDB | 0.545641 | RDF085m | -0.627684 |
| R6u+ | 0.502103 | E1p | -0.620146 |
| GGI9 | 0.49145 | Mor25m | -0.603496 |
| O-058 | 0.484375 | R2v+ | -0.543559 |
| MATS6p | 0.478762 | RCI | -0.471169 |
| JGI4 | 0.464782 | JGI10 | -0.450297 |

| | | | |
|---|---|---|---|
| R5m+ | 0.449641 | BIC3 | -0.433845 |
| Mor21u | 0.418468 | E3s | -0.433083 |
| SPAM | 0.390473 | L3s | -0.400674 |
| RDF145u | 0.377485 | Dm | -0.344892 |
| R2u+ | 0.347814 | Mor06v | -0.305683 |
| E2e | 0.324195 | R3v+ | -0.299862 |
| Mor08u | 0.264495 | Mor29v | -0.297368 |
| Mor22e | 0.23801 | Mor10p | -0.297121 |
| Mor19u | 0.235263 | RTp+ | -0.294698 |
| PW5 | 0.218135 | R5u+ | -0.288912 |
| RDF045m | 0.17446 | JGI1 | -0.286692 |
| Mor22u | 0.165338 | RDF070m | -0.285107 |
| Mor04p | 0.165232 | RTv+ | -0.277181 |
| TOS | 0.164877 | EEig12x | -0.267931 |
| Mor23m | 0.152134 | Mor10v | -0.236662 |
| JGI8 | 0.137205 | RDF090e | -0.226201 |
| Mor16m | 0.13109 | Mor13u | -0.21849 |
| H7u | 0.123767 | BELv5 | -0.218087 |
| MATS1m | 0.114605 | Mor10u | -0.205804 |
| H6m | 0.106284 | RDF055m | -0.202087 |
| nCconj | 0.105095 | R8e+ | -0.182816 |
| GATS5m | 0.092177 | R6v+ | -0.168503 |
| GATS5p | 0.089884 | GATS8e | -0.142111 |
| SRW09 | 0.088207 | HOMA | -0.127141 |
| Mor14e | 0.072911 | BELp5 | -0.123678 |
| R5e+ | 0.071676 | RDF095m | -0.122296 |
| G3p | 0.071423 | R1m+ | -0.121098 |
| RDF120m | 0.071132 | RDF050v | -0.113444 |
| Mor24u | 0.05863 | L2p | -0.111242 |
| MATS8m | 0.057825 | LAI | -0.106511 |

| | | | |
|---|---|---|---|
| JGI7 | 0.050859 | R6p+ | -0.105849 |
| nCs | 0.050296 | H8u | -0.098917 |
| Mor28m | 0.04959 | RDF090m | -0.096145 |
| Mor31m | 0.048422 | Hypnotic-50 | -0.092558 |
| RDF115u | 0.046823 | R2p+ | -0.083373 |
| RDF110u | 0.044665 | RDF155v | -0.079466 |
| RDF145e | 0.042256 | Mor30e | -0.071162 |
| C-025 | 0.039671 | Mor02v | -0.064702 |
| R6e+ | 0.037562 | BELm5 | -0.063507 |
| G3v | 0.037239 | BELp8 | -0.059291 |
| MATS8p | 0.036776 | Ku | -0.059057 |
| P2s | 0.035681 | DISPp | -0.055435 |
| RDF125m | 0.034567 | MATS8e | -0.050401 |
| Mor12u | 0.033794 | L2v | -0.050032 |
| Mor28u | 0.033505 | R8u | -0.048706 |
| G2s | 0.031251 | AROM | -0.044368 |
| Mor18m | 0.030071 | R3m+ | -0.0398 |
| Mor24v | 0.026671 | Psychotic-80 | -0.038048 |
| Psychotic-50 | 0.024805 | RTm+ | -0.034295 |
| GNF | 0.023216 | DISPe | -0.033192 |
| Mor03e | 0.021333 | Mor30m | -0.032476 |
| Mor05u | 0.021315 | GATS8p | -0.032221 |
| TNN | 0.020443 | SPH | -0.031749 |
| Mor08e | 0.020156 | Mor28v | -0.030508 |
| RDF155m | 0.019378 | RDF070e | -0.025199 |
| R7m+ | 0.015938 | MATS8v | -0.022534 |
| FDI | 0.015222 | E1m | -0.019097 |
| RDF150m | 0.015163 | PJI3 | -0.015838 |
| R8m+ | 0.01308 | R4m+ | -0.015492 |
| Mor16e | 0.012181 | P1u | -0.015364 |

| | | | |
|---|---|---|---|
| Mor08p | 0.010846 | Mor12m | -0.014549 |
| RDF140v | 0.01021 | G2m | -0.011805 |
| TNF | 0.006238 | RDF055v | -0.010748 |
| C-039 | 0.004455 | Mor21m | -0.010587 |
| R8p+ | 0.004017 | RDF155p | -0.0087 |
| Mor22p | 0.00348 | H8m | -0.006323 |
| RDF105m | 0.00193 | Mor09u | -0.004866 |
| DISPv | 0.001492 | RDF140u | -0.003518 |
| C-002 | 0.000152 | GOF | -0.001866 |
| | | G2v | -0.001213 |
| | | GATS8v | -0.000789 |
| | | TOF | -0.000024 |



\end{document}